\documentclass[a4paper,11pt]{article}
\pdfoutput=1 % if your are submitting a pdflatex (i.e. if you have
\usepackage{jheppub} % for details on the use of the package, please
\usepackage[T1]{fontenc} % if needed

\RequirePackage[displaymath]{lineno} % Display line numbers
\usepackage{epstopdf}
\usepackage{epsfig}
\usepackage{graphicx}% Include figure files
\usepackage{dcolumn}% Align table columns on decimal point
\usepackage{bm}% bold math.pdf
\usepackage{ltablex,booktabs}
\usepackage{overpic}
\usepackage{subfigure}
\usepackage{float}
\usepackage{color}
\usepackage{amsmath}
\usepackage{mathcomp}
\usepackage{mathrsfs}
\usepackage{multirow}
\usepackage{makecell}
\usepackage{rotating}
\usepackage{amssymb}
\usepackage{gensymb}
\usepackage{amsmath}
\usepackage{tabularx}
\usepackage{threeparttable}
\begin{document}

%\linenumbers
\title{Measurement of $e^+e^-\to pK^-\bar{\Lambda}+c.c.$ cross sections between 4.009 GeV and 4.951 GeV}

\collaborationImg{\includegraphics[width=0.15\textwidth, angle=90]{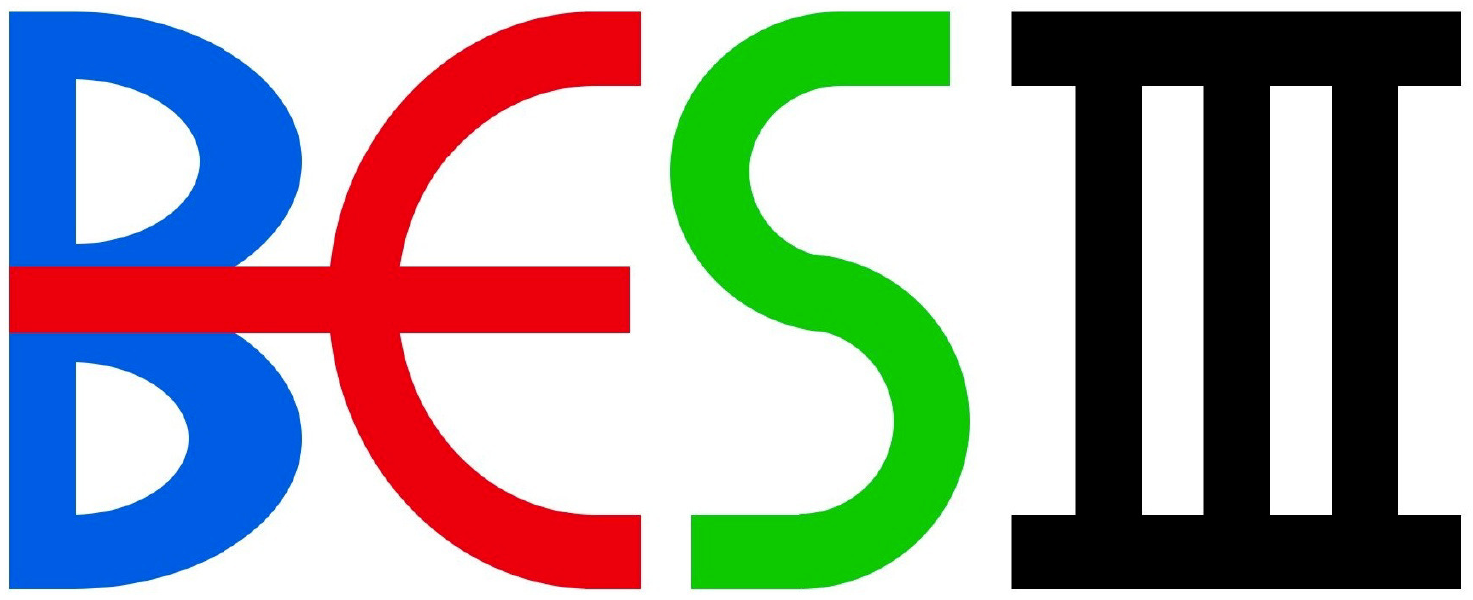}}
\collaboration{BESIII Collaboration}

%\input{author}

%% Saved at => 2023-01-13
\author{
M.~Ablikim$^{1}$, M.~N.~Achasov$^{13,b}$, P.~Adlarson$^{75}$, X.~C.~Ai$^{81}$, R.~Aliberti$^{36}$, A.~Amoroso$^{74A,74C}$, M.~R.~An$^{40}$, Q.~An$^{71,58}$, Y.~Bai$^{57}$, O.~Bakina$^{37}$
% ...
\\
\vspace{0.2cm}
(BESIII Collaboration)\\
\vspace{0.2cm} {\it
$^{1}$ Institute of High Energy Physics, Beijing 100049, People's Republic of China\\
% ...
$^{13}$ G.I. Budker Institute of Nuclear Physics SB RAS (BINP), Novosibirsk 630090, Russia\\
% ...
$^{36}$ Johannes Gutenberg University of Mainz, Johann-Joachim-Becher-Weg 45, D-55099 Mainz, Germany\\
$^{37}$ Joint Institute for Nuclear Research, 141980 Dubna, Moscow region, Russia\\
% ...
$^{40}$ Liaoning Normal University, Dalian 116029, People's Republic of China\\
% ...
$^{57}$ Southeast University, Nanjing 211100, People's Republic of China\\
$^{58}$ State Key Laboratory of Particle Detection and Electronics, Beijing 100049, Hefei 230026, People's Republic of China\\
% ...
$^{71}$ University of Science and Technology of China, Hefei 230026, People's Republic of China\\
% ...
$^{74}$ University of Turin and INFN, (A)University of Turin, I-10125, Turin, Italy; (B)University of Eastern Piedmont, I-15121, Alessandria, Italy; (C)INFN, I-10125, Turin, Italy\\
$^{75}$ Uppsala University, Box 516, SE-75120 Uppsala, Sweden\\
% ...
$^{81}$ Zhengzhou University, Zhengzhou 450001, People's Republic of China\\

\vspace{0.2cm}
% ...
$^{b}$ Also at the Novosibirsk State University, Novosibirsk, 630090, Russia\\
% ...
}
%% ends here %%
}

\abstract{Using $e^+e^-$ collision datasets corresponding to total integrated luminosity of 21.7 fb$^{-1}$ collected with the BESIII detector at the BEPCII collider at center-of-mass energies ranging from 4.009 GeV to 4.951 GeV, the energy-dependent cross sections of $e^+e^-\to pK^-\bar{\Lambda}+c.c.$ are measured for the first time. By fitting these energy-dependent cross sections, we search for the excited $\psi$ states $\psi(4160)$ and $\psi(4415)$, and the vector charmonium-like states $\psi(4230)$, $\psi(4360)$, and $\psi(4660)$. No evidence for these is observed and the upper limits on the branching fractions of these states decaying into $pK^-\bar \Lambda+c.c.$ are set at the 90\% confidence level.
}

%\arxivnumber{}

%\begin{document}
\maketitle
\flushbottom

%------------------------------------------------------------------------------
\section{Introduction}

In the last two decades, a large number of charmonium-like vector states $\psi$ have been discovered~\cite{PDG2022} in the hidden or open charm final states. The $\psi(4260)$ was first observed by the BaBar Collaboration in the process of $e^+e^-\to\gamma_{\rm ISR}\pi^+\pi^-J/\psi$ ~\cite{BABAR_exp:Y4260}, where ISR denotes initial state radiation. The  $\psi(4360)$ and $\psi(4660)$ were found by the Belle and BaBar Collaborations in the $\pi^+\pi^-\psi(2S)$ final states~\cite{BELLE_exp:Y4360_Y4660,BABAR_exp:Y4360}. Later, a precise study on the process $e^+e^-\to \pi^+\pi^-J/\psi$ by the BESIII Collaboration revealed two structures with masses of $4222.0\pm3.1\pm1.4$ MeV/$c^2$ and $4320.0\pm10.4\pm7.0$ MeV/$c^2$ in the $\psi(4260)$ region~\cite{BESIII_exp:pipiJpsi}. The former one, renamed as $Y(4230)$~\cite{PDG2022}, was further confirmed by the BESIII Collaboration in the decay channels $e^+e^-\to\omega\chi_{c0}$~\cite{BESIII_exp:omegaChic0}, $e^+e^-\to\pi^+\pi^-h_{c}$~\cite{BESIII_exp:pipiHc}, $e^+e^-\to\pi^+D^0D^{*-}$~\cite{BESIII_exp:piDDstar}, $e^+e^-\to\eta J/\psi$~\cite{BESIII_exp:etaJpsi}, $e^+e^-\to\pi^+\pi^+\psi(3686)$~\cite{BESIII_exp:pipipsip}, and $e^+e^-\to\pi^+D^{*0}D^{*-}$~\cite{BESIII_exp:DstarDstarpi}.

Since some properties of these states cannot be explained by the conventional charmonium model, they are usually regarded as candidates for exotic states, such as hybrids, tetraquarks, and molecules~\cite{Ystate_th1,Ystate_th2,Ystate_th3}.  Searching for the $Y$ states decaying into light hadron final states will help to understand the nature of $Y$ states and investigate the mechanism of quantum chromodynamics at low energies. Although several processes with light hadron final states, such as $e^+e^-\to p\bar{p}\pi^0$~\cite{BES_exp:ee2pppi0}, $e^+e^-\to p\bar{p}\eta(\omega)$~\cite{BES_exp:ee2ppeta_omega}, $e^+e^-\to p\bar{n}K^0_{S}K^-$~\cite{BES_exp:ee2pksnk}, $e^+e^-\to K^0_SK^\pm\pi^{\mp}\pi^0(\eta)$~\cite{BES_exp:ee2kkpipi_eta}, $e^+e^-\to\omega\pi^+\pi^-$~\cite{BESIII_exp:omegapipi}, have been studied, no significant charmonium-like structures were found in these. Consequently, further exploration of $e^+e^-$ decaying into other light hadrons is highly desirable to probe the nature of these charmonium-like states.

In this analysis, the cross sections of the process $e^+e^-\to pK^-\bar{\Lambda}+c.c.$ are measured by analyzing 21.7 fb$^{-1}$ of $e^+e^-$ collision data taken at center-of-mass energies ($\sqrt{s}$) ranging from 4.009 GeV to 4.951 GeV. The vector charmonium(-like) states, $\psi(4160)$, $\psi(4230)$, $\psi(4360)$, $\psi(4415)$, and $\psi(4660)$ are investigated by fitting the obtained energy-dependent cross sections. Throughout the paper, the charged-conjugation mode
is always implied, unless explicitly stated.

%------------------------------------------------------------------------------
\section{Detector and data sets}

The BESIII detector is a magnetic spectrometer~\cite{Ablikim:2009aa}
located at the Beijing Electron Positron Collider~(BEPCII)~\cite{Yu:IPAC2016-TUYA01}.
The cylindrical core of the BESIII detector consists of a main drift chamber filled with helium-based gas~(MDC), a plastic scintillator time-of-flight
system~(TOF), and a CsI(Tl) electromagnetic calorimeter~(EMC), which are all
enclosed in a superconducting solenoidal magnet providing a 1.0~T magnetic
field. The flux-return yoke is instrumented with resistive plate
chambers arranged in 9 layers in the barrel and 8 layers in the endcaps for muon identification. The
acceptance of charged particles and photons is 93\% of $4\pi$ solid angle.
The charged-particle momentum resolution at 1.0~GeV/$c$ is $0.5\%$, and the
specific energy loss resolution is $6\%$ for the electrons from
Bhabha scattering. The EMC measures photon energies with a resolution of
$2.5\%$~($5\%$) at $1$~GeV in the barrel (end cap) region. The time resolution
of the TOF barrel part is 68~ps, while that of the end cap part is 110~ps. The end cap TOF system was upgraded in 2015 with multi-gap resistive plate chamber technology, providing a time resolution of 60 ps~\cite{TOF_update1,TOF_update2,TOF_update3}. All of those are enclosed in a superconducting solenoidal magnet providing a 1.0 T magnetic field~\cite{magnet}.

%------------------------------------------------------------------------------
The data samples used in this analysis were collected by the BESIII detector at 37 energy points between 4.009 GeV and 4.951 GeV. The center-of-mass energies and the corresponding integrated luminosities~\cite{cited_luminosity1,cited_luminosity2} at various energy points are shown in Table \ref{tab:Summary_tab_cross_section}. Simulated samples produced with the {\sc geant4}-based~\cite{GEANT4}
Monte-Carlo~(MC) software, which includes the geometric description of
the BESIII detector and the detector response, are used to determine
the detection efficiencies and to estimate the background levels. The simulation includes the beam energy spread and ISR in the $e^+e^-$ annihilations modeled with the generator {\sc kkmc}~\cite{KKMC}. Inclusive MC simulation samples generated at $\sqrt{s}=4.178$ GeV with 40 times the luminosity of the data sample are used to analyze the possible background contributions. They consist of open charm production processes, ISR production of vector charmonium or charmonium-like states, and continuum processes ($e^+e^-\to q\bar{q},q=u,d,s$). The open charm production processes are generated using {\sc conexc} \cite{conexc}, and the ISR production is incorporated in {\sc kkmc}~\cite{KKMC}. The known decay states are modeled with {\sc beseventgen}~\cite{EVTGEN1,EVTGEN2} using branching fractions taken from the Particle Data Group~(PDG)~\cite{PDG2022} and the remaining unknown decays from the charmonium states are modeled with {\sc lundcharm}~\cite{LUNDCHARM1,LUNDCHARM2}. Final state radiation from charged final state particles is incorporated with the {\sc photos}~\cite{PHOTOS} package. The signal MC samples of $e^+e^-\to pK^-\bar{\Lambda}$ are generated by using the amplitude model with parameters fixed to the amplitude analysis results~\cite{BESIII:pkl_X2085}.

\begin{table*}
	\setlength\tabcolsep{9pt}
	\centering
	\caption{The Born cross sections of the $e^+e^-\to pK^-\bar{\Lambda}+c.c.$ process at various energy points. Here $\mathcal{L}_{\rm int}$ is the integrated luminosity, $N_{\rm sig}$ is the number of signal events from the fit to the $M_{p\pi^-}$ distribution, $\epsilon$ the detection efficiency, $(1+\delta)$ the radiative correction factor including vacuum polarization effect, $\frac{1}{1-\Pi^2}$ the vacuum polarization factor, and $\sigma^{\rm B}$ the Born cross section. The first uncertainties are statistical and the second systematic.}
	\begin{tabular*}{\textwidth}{lllllll}
		\hline
		\hline
		\makecell[c]{$\sqrt{s}$ (GeV)} & \makecell[c]{$\mathcal{L}_{\rm int}$ (pb$^{-1}$)} &  \makecell[c]{$N_{\rm sig}$}& \makecell[c]{$\epsilon$~(\%)} & \makecell[c]{($1+\delta$)} & $\frac{1}{1-\Pi^2}$ & \makecell[c]{$\sigma^{\rm B}$~(pb)}\\ \hline
		\makecell[c]{4.009}  & \makecell[c]{482.0$\pm$4.7} & \makecell[c]{$754 \pm 28$} &  \makecell[c]{20.2}  &  \makecell[c]{1.16} & \makecell{1.04}   &\makecell[c]{$10.1\pm0.4\pm0.4$} \\
		\makecell[c]{4.129}  & \makecell[c]{401.5$\pm$2.6} & \makecell[c]{$507 \pm 24$} &  \makecell[c]{20.1}  &  \makecell[c]{1.19} & \makecell{1.05}  & \makecell[c]{$7.9\pm0.4\pm0.3$}  \\
		\makecell[c]{4.157}  & \makecell[c]{408.7$\pm$2.6} & \makecell[c]{$463 \pm 23$} &  \makecell[c]{20.4}  &  \makecell[c]{1.20} & \makecell{1.05}  & \makecell[c]{$6.9\pm0.3\pm0.3$}  \\
		\makecell[c]{4.178}  & \makecell[c]{3189.0$\pm$31.9}& \makecell[c]{$4025\pm 67$} &  \makecell[c]{20.3}  & \makecell[c]{1.20} & \makecell{1.05}   &\makecell[c]{$7.7\pm0.1\pm0.3$}  \\
		\makecell[c]{4.189}  & \makecell[c]{526.7$\pm$2.2} & \makecell[c]{$658 \pm 27$} &  \makecell[c]{20.3}  &  \makecell[c]{1.21} & \makecell{1.06}  & \makecell[c]{$7.6\pm0.3\pm0.4$}  \\
		\makecell[c]{4.199}  & \makecell[c]{526.0$\pm$2.5} & \makecell[c]{$639 \pm 26$} &  \makecell[c]{20.4}  &  \makecell[c]{1.21} & \makecell{1.06}  & \makecell[c]{$7.3\pm0.3\pm0.3$}  \\
		\makecell[c]{4.209}  & \makecell[c]{517.1$\pm$1.8} & \makecell[c]{$618 \pm 25$} &  \makecell[c]{20.1}  &  \makecell[c]{1.21} & \makecell{1.06}  & \makecell[c]{$7.3\pm0.3\pm0.3$}  \\
		\makecell[c]{4.219}  & \makecell[c]{514.6$\pm$1.8} & \makecell[c]{$639 \pm 27$} &  \makecell[c]{19.4}  &  \makecell[c]{1.22} & \makecell{1.06}  & \makecell[c]{$7.8\pm0.3\pm0.4$}  \\
		\makecell[c]{4.226}  & \makecell[c]{1100.9$\pm$7.0}& \makecell[c]{$1247\pm 38$} &  \makecell[c]{20.4}  &  \makecell[c]{1.22} & \makecell{1.06}  & \makecell[c]{$6.8\pm0.2\pm0.3$}  \\
		\makecell[c]{4.236}  & \makecell[c]{530.3$\pm$2.4} & \makecell[c]{$630 \pm 26$} &  \makecell[c]{19.8}  &  \makecell[c]{1.22} & \makecell{1.06}  & \makecell[c]{$7.3\pm0.3\pm0.3$}  \\
		\makecell[c]{4.244}  & \makecell[c]{538.1$\pm$2.7} & \makecell[c]{$570 \pm 25$} &  \makecell[c]{20.3}  &  \makecell[c]{1.23} & \makecell{1.06}  & \makecell[c]{$6.3\pm0.3\pm0.5$}  \\
		\makecell[c]{4.258}  & \makecell[c]{828.4$\pm$5.5} & \makecell[c]{$932 \pm 32$} &  \makecell[c]{20.5}  &  \makecell[c]{1.23} & \makecell{1.05}  & \makecell[c]{$6.6\pm0.2\pm0.7$}  \\
		\makecell[c]{4.267}  & \makecell[c]{531.1$\pm$3.1} & \makecell[c]{$589 \pm 25$} &  \makecell[c]{20.4}  &  \makecell[c]{1.23} & \makecell{1.05}  & \makecell[c]{$6.6\pm0.3\pm0.5$}  \\
		\makecell[c]{4.278}  & \makecell[c]{175.7$\pm$1.0} & \makecell[c]{$197 \pm 15$} &  \makecell[c]{20.1}  &  \makecell[c]{1.24} & \makecell{1.05}  & \makecell[c]{$6.7\pm0.5\pm0.6$}  \\
		\makecell[c]{4.288}  & \makecell[c]{502.4$\pm$3.3} & \makecell[c]{$496 \pm 24$} &  \makecell[c]{19.8}  &  \makecell[c]{1.24} & \makecell{1.05}  & \makecell[c]{$6.0\pm0.3\pm0.4$}  \\
		\makecell[c]{4.312}  & \makecell[c]{501.2$\pm$3.3} & \makecell[c]{$566 \pm 23$} &  \makecell[c]{19.8}  &  \makecell[c]{1.25} & \makecell{1.05}  & \makecell[c]{$6.8\pm0.3\pm0.4$}  \\
		\makecell[c]{4.337}  & \makecell[c]{505.0$\pm$3.4} & \makecell[c]{$520 \pm 24$} &  \makecell[c]{20.1}  &  \makecell[c]{1.26} & \makecell{1.05}  & \makecell[c]{$6.1\pm0.3\pm0.3$}  \\
		\makecell[c]{4.358}  & \makecell[c]{543.9$\pm$3.6} & \makecell[c]{$550 \pm 25$} &  \makecell[c]{21.0}  &  \makecell[c]{1.26} & \makecell{1.05}  & \makecell[c]{$5.7\pm0.3\pm0.3$}  \\
		\makecell[c]{4.377}  & \makecell[c]{522.7$\pm$3.5} & \makecell[c]{$529 \pm 24$} &  \makecell[c]{20.3}  &  \makecell[c]{1.27} & \makecell{1.05}  & \makecell[c]{$5.9\pm0.3\pm0.3$}  \\
		\makecell[c]{4.397}  & \makecell[c]{507.8$\pm$3.4} & \makecell[c]{$449 \pm 23$} &  \makecell[c]{20.3}  &  \makecell[c]{1.27} & \makecell{1.05}  & \makecell[c]{$5.1\pm0.3\pm0.3$}  \\
		\makecell[c]{4.416}  & \makecell[c]{1090.7$\pm$7.2}& \makecell[c]{$1030\pm 33$} &  \makecell[c]{20.4}  &  \makecell[c]{1.28} & \makecell{1.05}  & \makecell[c]{$5.4\pm0.2\pm0.2$}  \\
		\makecell[c]{4.436}  & \makecell[c]{569.9$\pm$3.8} & \makecell[c]{$523 \pm 24$} &  \makecell[c]{20.2}  &  \makecell[c]{1.29} & \makecell{1.05}  & \makecell[c]{$5.3\pm0.2\pm0.3$}  \\
		\makecell[c]{4.467}  & \makecell[c]{111.1$\pm$0.8} & \makecell[c]{$90  \pm 10$} &  \makecell[c]{20.2}  &  \makecell[c]{1.30} & \makecell{1.05}  & \makecell[c]{$4.6\pm0.5\pm0.3$}  \\
		\makecell[c]{4.527}  & \makecell[c]{112.1$\pm$0.8} & \makecell[c]{$87  \pm 9 $} &  \makecell[c]{19.8}  &  \makecell[c]{1.32} & \makecell{1.05}  & \makecell[c]{$4.4\pm0.5\pm0.2$}  \\
		\makecell[c]{4.600}  & \makecell[c]{586.9$\pm$4.0} & \makecell[c]{$416 \pm 21$} &  \makecell[c]{21.8}  &  \makecell[c]{1.35} & \makecell{1.05}  & \makecell[c]{$3.6\pm0.2\pm0.2$}  \\
		\makecell[c]{4.612}  & \makecell[c]{103.7$\pm$0.6} & \makecell[c]{$81  \pm 9 $} &  \makecell[c]{21.1}  &  \makecell[c]{1.36} & \makecell{1.05}  & \makecell[c]{$4.0\pm0.5\pm0.3$}  \\
		\makecell[c]{4.628}  & \makecell[c]{521.5$\pm$2.8} & \makecell[c]{$357 \pm 20$} &  \makecell[c]{20.7}  &  \makecell[c]{1.38} & \makecell{1.05}  & \makecell[c]{$3.6\pm0.2\pm0.2$}  \\
		\makecell[c]{4.641}  & \makecell[c]{551.7$\pm$3.0} & \makecell[c]{$375 \pm 20$} &  \makecell[c]{20.7}  &  \makecell[c]{1.37} & \makecell{1.05}  & \makecell[c]{$3.6\pm0.2\pm0.2$}  \\
		\makecell[c]{4.661}  & \makecell[c]{529.4$\pm$2.8} & \makecell[c]{$351 \pm 20$} &  \makecell[c]{20.7}  &  \makecell[c]{1.38} & \makecell{1.05}  & \makecell[c]{$3.5\pm0.2\pm0.2$}  \\
		\makecell[c]{4.682}  & \makecell[c]{1667.4$\pm$8.9}& \makecell[c]{$1012\pm 34$} &  \makecell[c]{20.6}  &  \makecell[c]{1.39} & \makecell{1.05}  & \makecell[c]{$3.2\pm0.1\pm0.2$}  \\
		\makecell[c]{4.699}  & \makecell[c]{535.5$\pm$2.9} & \makecell[c]{$362 \pm 20$} &  \makecell[c]{20.2}  &  \makecell[c]{1.39} & \makecell{1.05}  & \makecell[c]{$3.6\pm0.2\pm0.2$}  \\
		\makecell[c]{4.740}  & \makecell[c]{163.9$\pm$0.9} & \makecell[c]{$115 \pm 11$} &  \makecell[c]{20.9}  &  \makecell[c]{1.40} & \makecell{1.05}  & \makecell[c]{$3.5\pm0.3\pm0.3$}  \\
		\makecell[c]{4.750}  & \makecell[c]{366.6$\pm$2.0} & \makecell[c]{$248 \pm 17$} &  \makecell[c]{20.8}  &  \makecell[c]{1.41} & \makecell{1.05}  & \makecell[c]{$3.4\pm0.2\pm0.2$}  \\
		\makecell[c]{4.781}  & \makecell[c]{511.5$\pm$2.7} & \makecell[c]{$288 \pm 19$} &  \makecell[c]{20.5}  &  \makecell[c]{1.42} & \makecell{1.06}  & \makecell[c]{$2.9\pm0.2\pm0.2$}  \\
		\makecell[c]{4.843}  & \makecell[c]{525.1$\pm$2.8} & \makecell[c]{$235 \pm 17$} &  \makecell[c]{20.0}  &  \makecell[c]{1.47} & \makecell{1.06}  & \makecell[c]{$2.3\pm0.2\pm0.2$}  \\
		\makecell[c]{4.918}  & \makecell[c]{207.8$\pm$1.1} & \makecell[c]{$127 \pm 11$} &  \makecell[c]{19.1}  &  \makecell[c]{1.55} & \makecell{1.06}  & \makecell[c]{$3.1\pm0.3\pm0.2$}  \\
		\makecell[c]{4.951}  & \makecell[c]{159.3$\pm$0.9} & \makecell[c]{$63  \pm 9 $} &  \makecell[c]{18.5}  &  \makecell[c]{1.59} & \makecell{1.06}  & \makecell[c]{$2.0\pm0.3\pm0.2$}  \\
		\hline
		\hline
	\end{tabular*}
	\label{tab:Summary_tab_cross_section}
\end{table*}

\section{Event selection and background analysis}
\label{chap:event_selection}

To select candidates for $e^+e^-\to pK^-\bar{\Lambda}$, the $\Lambda$ candidates are reconstructed with the charged decay mode, and hence there are only four charged tracks $p\bar{p}\pi^\pm K^\mp$ in the final states. All charged tracks are required to satisfy $|d_z|<$ 20 cm and $|\cos\theta|<$ 0.93. Here, $|d_z|$ is the coordinate of the charged particle production point along the beam axis and $\theta$ is the polar angle of the charged track. For each event, four charged tracks with zero net charge are required. Due to the high momenta of final charged tracks, particle identification is not required.

To reconstruct candidates for $\Lambda$, all possible opposite charged track pairs will be assigned as $p\pi^-$. The $p\pi^-$ trajectories are constrained to originate from a common vertex by applying a vertex fit, and the $\chi^2$ of the vertex fit is required to be less than 100. The $\Lambda$ candidate is constrained in a secondary vertex fit to originate from the interaction point. The decay length of the $\Lambda$ candidate must be greater than twice the vertex resolution.  The invariant mass of the $p\pi^-$ combination is required to be within 1.10 $<M_{p\pi^-}<$ 1.13 GeV/$c^2$ to suppress wrong assignments.  Exactly one $\Lambda$
candidate per event is required to satisfy the selection criteria. The other two charged tracks are assigned according to their charges as proton and kaon not from $\Lambda$ decays. To ensure that the two charged tracks originate from the interaction point, they are imposed with additional requirements of $|d_z|<$ 10 cm and $|d_r|<$ 1 cm. Here, $|d_r|$ is the distance between the charged track production point and the beam axis in the plane perpendicular to the beam axis.

To further suppress background and improve track momentum resolution, a four-momentum constraint~(4C) kinematic fit is imposed on the initial $e^+e^-$ beam energy under the hypothesis of $e^+e^-\to pK^-\bar{\Lambda}$ The $\chi^2$ of the kinematic fit is required to be less than 100. To reduce the number of background events caused by spurious kaons originating from Bhabha scattering events interacting with the materials in the detector, the events with $|\cos\theta_K|>0.83$ are vetoed, where $\theta_K$ is the polar angle of the charged kaon.

After applying all of the above selection criteria, studies of the inclusive MC sample indicate that the total background fraction is only 1.8\%. The background contributions are categorized into the peaking background, such as $e^+e^-\to \pi^+\Lambda\bar{\Sigma}^-$, and non-peaking background, such as $e^+e^-\to\rho^0 p\bar{p}$. Given that the peaking background fraction is lower than $1.0$\%, it is neglected in the further analysis and will be considered as one source of systematic uncertainties.

\section{Cross section measurement}
\label{chap:BFMEASUREMENT}

\begin{figure}[htbp]
	\centering
	\includegraphics[width=0.6\columnwidth]{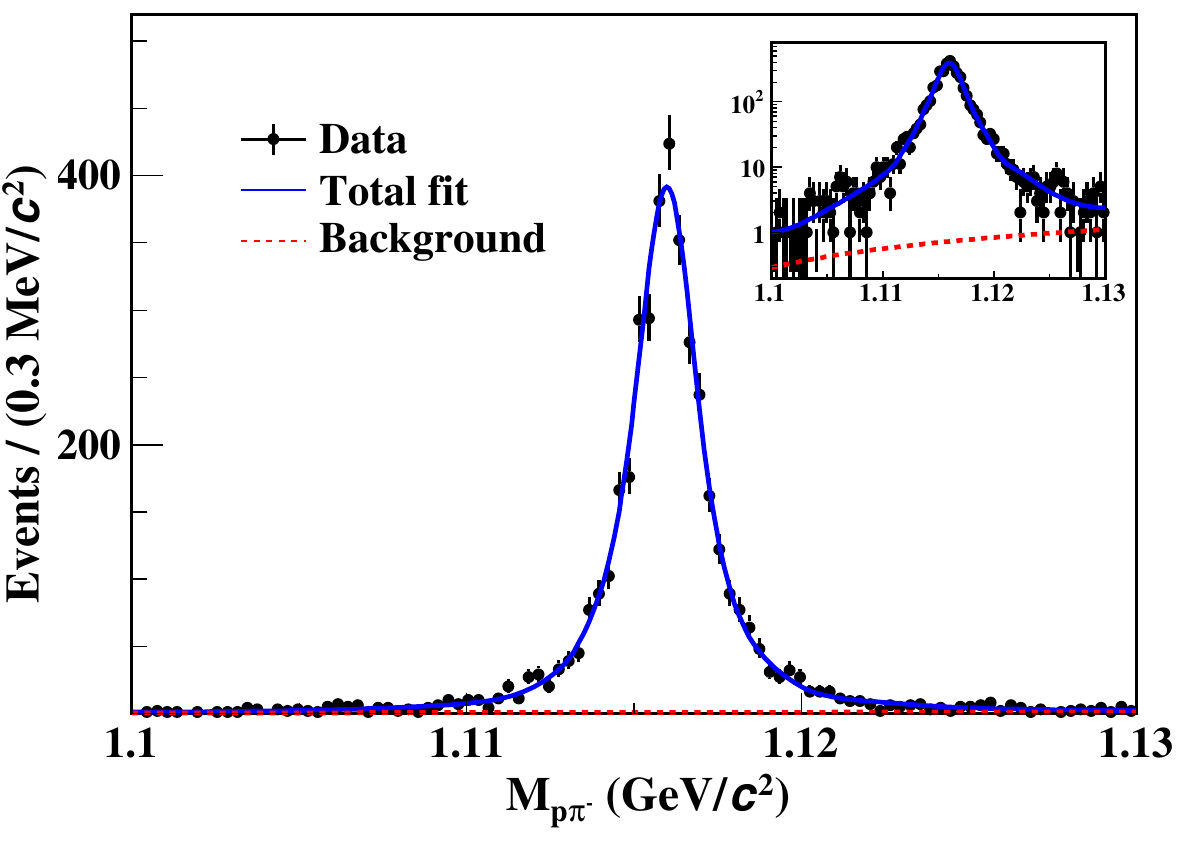}
	\caption{Fit result to the $p\pi^-$ invariant mass distribution of $e^+e^-\to pK^-\bar{\Lambda}+c.c.$. The points with errors are data at $\sqrt s=4.178$ GeV, the blue curve is the fit result, and the red dashed curve is the background contribution.
		\label{fig:fitppi}}
\end{figure}

To obtain the $e^+e^-\to pK^-\bar{\Lambda}$ signal yield, an un-binned maximum likelihood fit is performed to the invariant mass spectrum of $M_{p\pi^-}$. The signal is described with the MC-determined shape convolved with a Gaussian function to consider the difference between data and MC simulation. The background shape is parametrized as a linear function. As an example, Fig.~\ref{fig:fitppi} shows the fit result of the accepted candidates in data at $\sqrt s=4.178$ GeV, and all the signal yields~($N_{\rm sig}$) at 37 energy points are listed in Table~\ref{tab:Summary_tab_cross_section}.

In order to determine the detection efficiency, the amplitude analysis is performed for energy points with signal yield greater than 700, which are exactly the same as in Ref.~\cite{BESIII:pkl_X2085}. The amplitudes for the sequential processes $e^+e^-\to\gamma^*\to X^+ K^- (X^+\to p\bar{\Lambda})$, $e^+e^-\to\gamma^*\to N^{*+} \bar{p}(N^{*+}\to K^+\bar{\Lambda})$, $e^+e^-\to\gamma^*\to \Lambda^{*} \bar{\Lambda}(\Lambda^{*}\to pK^-)$, and their charge conjugations, are constructed using the relativistic covariant tensor amplitude formalism~\cite{FDC}. These effective vertices $\Gamma$ are deduced from an effective Lagrangian by considering $C$- and $P$-parity invariance, Lorentz invariance, and $CPT$ invariance.

The amplitude of a process containing a specific resonance is written as
\begin{equation}
	\mathcal{A}_j=\epsilon_{*\alpha}(p_0,m)\bar u(p_1,\lambda_1)\Gamma_{1}^{\alpha\mu_1\mu_2...}\mathcal{P}_{\mu_1\mu_2...\nu_1\nu_2...}\Gamma_2^{\nu_1\nu_2...}\nonumber\times v(p_2,\lambda_2)BW(s),
\end{equation}
where $\epsilon^{*}$ is the $\gamma^*$ polarization vector; $u(p_1,\lambda_1)$ and $v(p_2,\lambda_2)$ are the free Dirac spinors for proton and $\bar\Lambda$, respectively; $\Gamma_1$ and $\Gamma_2$ are the two strong interaction vertices describing the resonance couplings with $\gamma^*,~p,~K^{-}$, and $\bar{\Lambda}$, and $BW(s)$ is a Breit-Wigner function for an intermediate states with a spin projection operator $\mathcal{P}$. The complex coupling constants of the amplitudes are determined by an un-binned maximum likelihood fit using {\sc minuit}~\cite{Minuit}. The background contribution is estimated with the inclusive MC sample and subtracted from the likelihood. The baseline solution is determined at $\sqrt{s}=4.178$ GeV, which includes the $p\bar{\Lambda}$ threshold enhancement $X(2085)$~\cite{BESIII:pkl_X2085}, ${K^*_2}(1980),{K^*_4}(2045),K_2(2250),\Lambda(1520),\Lambda(1890),\Lambda(2350),$$N(1720)$, and $N(2570)$. Except for $X(2085)$, the resonance parameters are fixed to the respective world average values~\cite{PDG2022}.

\begin{figure}[htbp]
	\centering
	\includegraphics[width=\columnwidth]{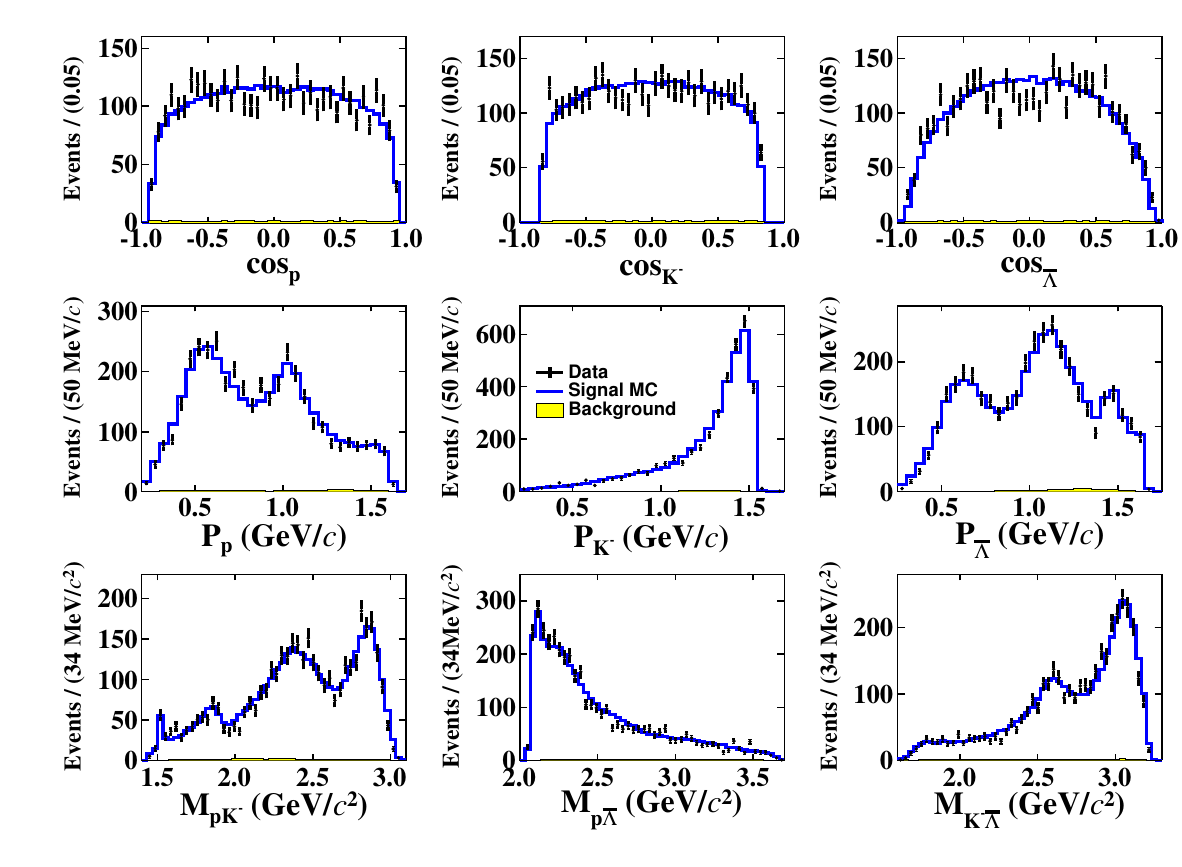}
	\caption{Distributions of $\cos\theta$, the momenta of daughter particles, and the invariant masses of all two-particle combinations at $\sqrt{s}=4.178$ GeV. The points with error bars are data, the blue solid curves are signal MC events, and the yellow hatched histogram is the normalized simulated background.}
	\label{fig:compare_toyMC}
\end{figure}

The signal MC samples of the other energy points are generated based on the amplitude analysis result of the nearby energy point. Figure~\ref{fig:compare_toyMC} shows the distributions of polar angles and momenta of final state particles, as well as the invariant mass spectra of all two-particle combinations of signal candidates in data and MC simulation at $\sqrt{s}=4.178$ GeV.

The Born cross section at a given center-of-mass energy is calculated as
\begin{equation}
\sigma^{\rm B}=\frac{N_{\rm sig}}{\mathcal{L}_{\rm int}\times\mathcal{B}\times\epsilon(1+\delta)\frac{1}{|1-\Pi|^2}},
\end{equation}
where $N_{\rm sig}$ is the fitted signal yield, $\mathcal{L}_{\rm int}$ is the integrated luminosity, $\mathcal{B}$ is the branching fraction of the $\Lambda$ charged decay~\cite{PDG2022}, $\epsilon$ is the detection efficiency, $(1+\delta)$ is the ISR correction factor, and $\frac{1}{|1-\Pi|^2}$ is the vacuum polarization factor \cite{conexc}.

\begin{figure}[htbp]
	\centering
	\includegraphics[width=0.6\columnwidth]{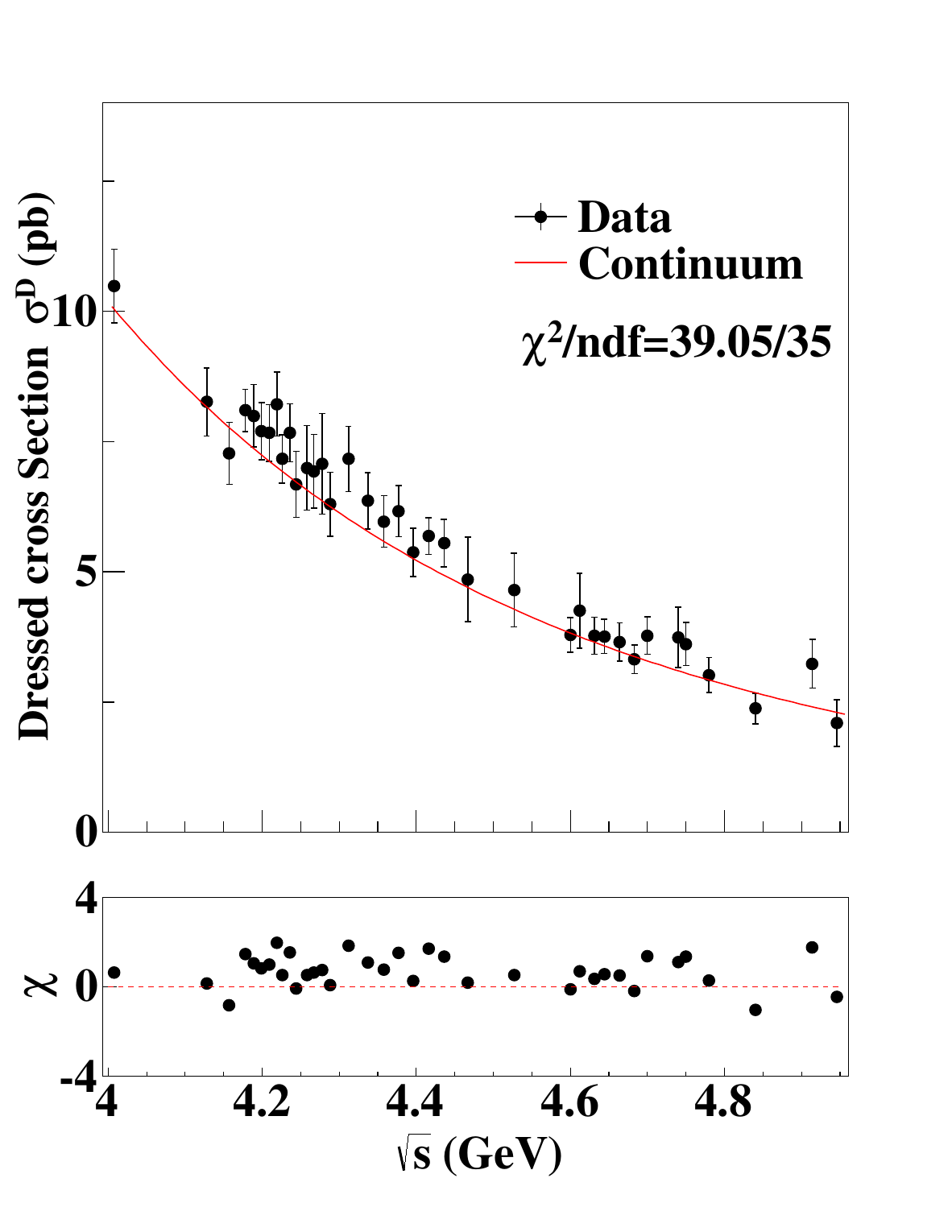}
	\caption{The energy-dependent dressed cross sections of the process $e^+e^-\to pK^-\bar{\Lambda}+c.c.$. The points with error bars are the measured values including both statistical and systematic uncertainties. The red solid curve is the fit result only with the continuum contribution $a/s^n$.  Due to the correlated systematic uncertainties, most of the $\chi$ values are positive by minimizing Eq.~\ref{eq:chisq_function}.}
	\label{fig:fit_continuum}
\end{figure}

To obtain the ISR correction factor, an iterative procedure is performed. First, a series of signal MC samples are generated for all energy points with a constant cross section. The cross sections are calculated based on the reconstruction efficiencies and correction factors obtained from the signal MC simulation. The line shape $\mathcal{P}^4(1-e^{-\Delta M/p_0})$ is used to describe the measured cross sections, where $\mathcal{P}^4$ is a forth-order polynomial with free parameters, $p_0$ is a free parameter, and $\Delta M=\sqrt{s}-2.645$ GeV since no signal event is observed with the data sample at $\sqrt{s}=2.645$ GeV, which is close to the mass threshold of $pK^-\bar{\Lambda}$. Then, the method introduced in Ref.~\cite{Iteration} is used to get the ISR correction factors and efficiencies, and a new series of cross sections is obtained. This procedure is repeated until the difference of $(1+\delta)\epsilon$ between two subsequent iterations is less than 0.1\%. The vacuum polarization factor $\frac{1}{|1-\Pi|^2}$ is obtained from {\sc conexc}~\cite{conexc}. Table~\ref{tab:Summary_tab_cross_section} summarizes the Born cross sections together with the relevant values used to determine them. The energy-dependent dressed cross sections of $e^+e^-\to pK^-\bar{\Lambda}$, defined as $\sigma^{\rm D}=\sigma^{\rm B}\times\frac{1}{|1-\Pi|^2}$, are shown in Fig.~\ref{fig:fit_continuum}.

%------------------------------------------------------------------------------
\section{Systematic uncertainty} \label{chap:SYSTEMATICS}

The systematic uncertainties on the cross section measurement include several sources, as summarized in Table~\ref{tab:sys_cross_sections}. They are estimated as described below.

\begin{table*}
	\setlength\tabcolsep{4.8pt}
	\centering
	\caption{Relative systematic uncertainties~(in \%) on the cross section measurement, which include luminosity~(Lum), tracking efficiency~(Trk), $\Lambda$ reconstruction efficiency~($\Lambda$ Rec), MC modeling~(MC), fit to $M_{p\pi^-}$~(Fit), 4C kinematic fit~(KMfit), $|\cos\theta_K|<$ 0.83~($\cos\theta$), correction factor ($1+\delta$), $\Lambda$ peaking background~(pBKG), and $\mathcal{B}(\Lambda\to p\pi^-)$ branching fraction~(BF).}
	\begin{tabular*}{\textwidth}{lcccccccccccc}
		\hline
		\hline
		$\sqrt{s}$ (GeV) & Lum & Trk & $\Lambda$ Rec & $N_{\rm trk}$ & MC & Fit &  KMfit & $\cos\theta$ & $1+\delta$ & pBKG & BF & Total \\
		\hline
4.009 & 1.0 & 3.0 & 2.0 & 0.5 & 0.7 & 0.8 & 0.3 & 0.9 & 0.5 & 1.0 & 0.8 & 4.3 \\
4.129 & 1.0 & 3.0 & 2.0 & 0.5 & 1.0 & 0.4 & 0.3 & 0.7 & 0.5 & 1.0 & 0.8 & 4.2 \\
4.157 & 1.0 & 3.0 & 2.0 & 0.5 & 0.9 & 0.4 & 0.5 & 0.8 & 0.5 & 1.0 & 0.8 & 4.3 \\
4.178 & 1.0 & 3.0 & 2.0 & 0.5 & 1.6 & 0.4 & 0.3 & 0.8 & 0.5 & 1.0 & 0.8 & 4.4 \\
4.189 & 1.0 & 3.0 & 2.0 & 0.5 & 1.3 & 2.0 & 0.4 & 0.8 & 0.5 & 1.0 & 0.8 & 4.8 \\
4.199 & 1.0 & 3.0 & 2.0 & 0.5 & 0.7 & 0.3 & 0.4 & 0.9 & 0.5 & 1.0 & 0.8 & 4.2 \\
4.209 & 1.0 & 3.0 & 2.0 & 0.5 & 0.3 & 0.6 & 0.3 & 0.8 & 0.5 & 1.0 & 0.8 & 4.2 \\
4.219 & 1.0 & 3.0 & 2.0 & 0.5 & 1.2 & 1.5 & 0.4 & 0.8 & 0.5 & 1.0 & 0.8 & 4.5 \\
4.226 & 1.0 & 3.0 & 2.0 & 0.5 & 1.7 & 1.8 & 0.2 & 0.9 & 0.5 & 1.0 & 0.8 & 4.8 \\
4.236 & 1.0 & 3.0 & 2.0 & 0.5 & 0.3 & 1.0 & 0.4 & 0.8 & 0.5 & 1.0 & 0.8 & 4.3 \\
4.244 & 1.0 & 3.0 & 2.0 & 0.5 & 5.7 & 1.5 & 0.5 & 0.9 & 0.5 & 1.0 & 0.8 & 7.2 \\
4.258 & 1.0 & 3.0 & 2.0 & 0.5 & 9.6 & 0.7 & 0.3 & 0.8 & 0.5 & 1.0 & 0.8 & 10.5 \\
4.267 & 1.0 & 3.0 & 2.0 & 0.5 & 7.1 & 0.6 & 0.4 & 0.8 & 0.5 & 1.0 & 0.8 & 8.2 \\
4.278 & 1.0 & 3.0 & 2.0 & 0.5 & 6.8 & 2.6 & 0.2 & 0.9 & 0.5 & 1.0 & 0.8 & 8.4 \\
4.288 & 1.0 & 3.0 & 2.0 & 0.5 & 5.5 & 1.2 & 0.3 & 0.9 & 0.5 & 1.0 & 0.8 & 7.0 \\
4.312 & 1.0 & 3.0 & 2.0 & 0.5 & 5.1 & 0.2 & 0.3 & 0.9 & 0.5 & 1.0 & 0.8 & 6.5 \\
4.337 & 1.0 & 3.0 & 2.0 & 0.5 & 3.4 & 0.7 & 0.3 & 0.8 & 0.5 & 1.0 & 0.8 & 5.4 \\
4.358 & 1.0 & 3.0 & 2.0 & 0.5 & 1.1 & 3.0 & 0.4 & 0.8 & 0.5 & 1.0 & 0.8 & 5.2 \\
4.377 & 1.0 & 3.0 & 2.0 & 0.5 & 2.3 & 0.4 & 0.3 & 0.8 & 0.5 & 1.0 & 0.8 & 4.7 \\
4.397 & 1.0 & 3.0 & 2.0 & 0.5 & 2.4 & 0.3 & 0.3 & 0.9 & 0.5 & 1.0 & 0.8 & 4.8 \\
4.416 & 1.0 & 3.0 & 2.0 & 0.5 & 0.2 & 0.6 & 0.3 & 0.9 & 0.5 & 1.0 & 0.8 & 4.2 \\
4.436 & 1.0 & 3.0 & 2.0 & 0.5 & 2.9 & 1.2 & 0.3 & 0.9 & 0.5 & 1.0 & 0.8 & 5.2 \\
4.467 & 1.0 & 3.0 & 2.0 & 0.5 & 0.9 & 4.0 & 0.4 & 0.8 & 0.5 & 1.0 & 0.8 & 5.8 \\
4.527 & 1.0 & 3.0 & 2.0 & 0.5 & 1.1 & 1.1 & 0.4 & 0.9 & 0.5 & 1.0 & 0.8 & 4.4 \\
4.600 & 1.0 & 3.0 & 2.0 & 0.5 & 3.1 & 0.3 & 0.3 & 0.8 & 0.6 & 1.0 & 0.8 & 5.2 \\
4.612 & 1.0 & 3.0 & 2.0 & 0.5 & 4.6 & 1.2 & 0.3 & 0.8 & 0.8 & 1.0 & 0.8 & 6.3 \\
4.628 & 1.0 & 3.0 & 2.0 & 0.5 & 2.4 & 0.6 & 0.6 & 0.8 & 1.4 & 1.0 & 0.8 & 5.0 \\
4.641 & 1.0 & 3.0 & 2.0 & 0.5 & 1.3 & 0.4 & 0.4 & 0.8 & 0.8 & 1.0 & 0.8 & 4.4 \\
4.661 & 1.0 & 3.0 & 2.0 & 0.5 & 3.9 & 1.0 & 0.3 & 0.9 & 0.8 & 1.0 & 0.8 & 5.8 \\
4.682 & 1.0 & 3.0 & 2.0 & 0.5 & 5.1 & 1.0 & 0.4 & 0.9 & 1.0 & 1.0 & 0.8 & 6.7 \\
4.699 & 1.0 & 3.0 & 2.0 & 0.5 & 3.1 & 1.1 & 0.4 & 0.9 & 0.7 & 1.0 & 0.8 & 5.3 \\
4.740 & 1.0 & 3.0 & 2.0 & 0.5 & 6.2 & 1.2 & 0.7 & 0.8 & 1.0 & 1.0 & 0.8 & 7.6 \\
4.750 & 1.0 & 3.0 & 2.0 & 0.5 & 4.2 & 1.5 & 0.3 & 0.9 & 1.1 & 1.0 & 0.8 & 6.2 \\
4.781 & 1.0 & 3.0 & 2.0 & 0.5 & 3.6 & 2.2 & 0.4 & 0.8 & 1.5 & 1.0 & 0.8 & 6.1 \\
4.843 & 1.0 & 3.0 & 2.0 & 0.5 & 5.2 & 1.3 & 0.4 & 0.9 & 1.3 & 1.0 & 0.8 & 6.9 \\
4.918 & 1.0 & 3.0 & 2.0 & 0.5 & 6.3 & 0.0 & 0.4 & 0.9 & 1.9 & 1.0 & 0.8 & 7.8 \\
4.951 & 1.0 & 3.0 & 2.0 & 0.5 & 4.4 & 3.5 & 0.4 & 0.8 & 3.3 & 1.0 & 0.8 & 7.7 \\
		\hline
		\hline
	\end{tabular*}
	\label{tab:sys_cross_sections}
\end{table*}

The integrated luminosity is measured using Bhabha scattering events, with uncertainty less than 1.0\%~\cite{cited_luminosity1,cited_luminosity2}. The uncertainty related to the tracking efficiency of kaons is estimated to be 1.0\% using a control sample of $e^+e^-\to K^+K^-\pi^+\pi^-$ and that of protons not from $\Lambda$ is estimated to be 2.0\% using a control sample of $e^+e^-\to p\pi^-\bar{p}\pi^+$. The total systematic uncertainty from
tracking is assigned as a linear sum of kaon and non-$\Lambda$ proton contributions. The systematic uncertainty due to the $\Lambda$ reconstruction efficiency including tracking efficiencies of the $p\pi^-$ pair, decay length requirement, mass window, vertex fit, and second vertex fit, is assigned as 2.0\% using the control sample of $J/\psi(\psi(3686))\to\Lambda\bar{\Lambda}$~\cite{Lambda_reconstruction_eff}. The systematic uncertainty related to the $N_{\rm trk}=4$ requirement, i.e. the number of charged tracks must be four, is estimated to be 0.5\% with a control sample of $J/\psi\to pK^-\bar{\Lambda}$ following Ref.~\cite{Ntrk_four_sys}.

The systematic uncertainty of the MC modeling is estimated with a new signal MC sample, in which all fitted complex coupling constants, quoted resonance parameters are smeared with their uncertainties. The difference between the detection efficiencies obtained with the new signal MC sample and the nominal one is taken as this uncertainty. The systematic uncertainty from the fit to the $M_{p\pi^-}$ spectrum is taken into account in two aspects. The uncertainty associated with the fit range is estimated by varying the fit range by 1~MeV. The uncertainty from the background shape is estimated by using a second-order polynomial function. For each of these two aspects, the maximum difference between the signal yields obtained with nominal and alternative background shapes is taken as the corresponding systematic uncertainty. Adding these two items in quadrature, we obtain the relevant systematic uncertainty.

The systematic uncertainty related to the 4C kinematic fit is estimated by comparing the detection efficiencies with and without the helix parameter correction~\cite{helix_sys}, taking the difference as the corresponding systematic uncertainty. The systematic uncertainty of the $|\cos\theta_K|<$ 0.83 requirement is estimated by varying the cut range by 0.01. The maximum difference between the detection efficiencies obtained with nominal and alternative cut ranges is taken to be the corresponding systematic uncertainty. The systematic uncertainty related to the correction factor is considered in two aspects. The uncertainty due to the theoretical uncertainty of the vacuum polarization factor is assigned to be 0.5\% ~\cite{conexc}. The uncertainty due to the line shape used in the iteration is estimated by varying all free parameters of the line shape within their statistical uncertainties. The distribution of $\sigma^{\rm B}$ obtained with the alternative parameter sets is fitted with a Gaussian function $(\mu_1,\sigma_1)$. The uncertainty is assigned to be $(|\mu_1-\mu_0|+\sigma_{1})/\mu_0\times100\%$, where $\mu_0$ is the nominal value. Adding these two items in quadrature gives the systematic uncertainty in the correction factor. The systematic uncertainty due to the $\Lambda$ peaking background is assigned as 1.0\% since the fraction of this background from the inclusive MC sample is found to be less than 1.0\%. The systematic uncertainty of the quoted branching fraction of $\Lambda\to p\pi^-$ is 0.8\%.

\section{Fit to the cross sections of $e^+e^-\to pK^-\bar{\Lambda}+c.c.$}
\label{chap:Fit_Xsec}

In order to search for possible charmless decays of charmonium(-like) states $\psi \to pK^-\bar{\Lambda}+c.c.$, we try two kinds of least chi-square fits to the dressed cross sections. The $\chi^2$ is constructed as

\begin{equation}
\chi^2= (\Delta\vec{\sigma})^T V^{-1} \Delta\vec{\sigma},
\label{eq:chisq_function}
\end{equation}
where $\Delta\vec{\sigma}_{i}=\sigma_{i}-\sigma^{\rm fit}_{i}(\vec{\theta})$ and $V$ is the covariance matrix. The $\sigma_{i}$ and $\sigma^{\rm fit}_{i}$ are the measured and fitted values for the cross section at the $i$-th energy point, respectively. The covariance matrix is constructed as $V_{ii}=V_{{\rm sta,}i}+V_{{\rm sys,}i}$ for diagonal elements and $V_{ij}=\sqrt{V_{{\rm corr-sys},i}\times V_{{\rm corr-sys},j}}$ for off-diagonal elements~($i\neq j$). Here, $V_{\rm corr-sys}$ includes the systematic uncertainties of integrated luminosity, tracking, $\Lambda$ reconstruction, and $\mathcal{B}(\Lambda\to p\pi^-)$.

In the first fit, the cross sections are assumed to result only from continuum production and to follow a relation of $a/s^n$. The fit result is shown in Fig.~\ref{fig:fit_continuum}, with the goodness-of-fit of $\chi^2/{\rm ndf}=39.05/35$, where both statistical and systematic uncertainties are included and ndf denotes the number of degrees of freedom. In the second fit, the cross section is modeled as a coherent sum of continuum production and resonant amplitudes, e.g., ${\sigma(\sqrt{s})=|a/s^n+BW(\sqrt{s})e^{i\phi}|^2}$, where
\begin{equation}
BW(\sqrt{s})=\frac{M}{\sqrt{s}}\frac{\sqrt{12\pi \Gamma_{\rm ee}\Gamma_{\rm tot}\mathcal{B}(\mathcal{R}\to pK^-\bar{\Lambda}+c.c.)}}{s-M^2+iM\Gamma_{\rm tot}}\sqrt{\frac{PS(\sqrt{s})}{PS(M)}}
\end{equation}
is used to describe charmonium-(like) states. Here, $M$, $\Gamma_{\rm tot}$, and $\Gamma_{\rm ee}$ are the mass, full width, and $e^+e^-$ partial width of the resonance $\mathcal{R}$, respectively. $\mathcal{B}(\mathcal{R}\to pK^-\bar{\Lambda}+c.c.)$ denotes the branching fraction of the decay $\mathcal{R}\to pK^-\bar{\Lambda}+c.c.$, $\phi$ is the relative phase between the continuum and resonance, and $\frac{PS(\sqrt{s})}{PS(M)}$ is the three-body phase space factor. In the second case, the values of $M$ and $\Gamma_{\rm tot}$ are fixed to the PDG values~\cite{PDG2022}. Several well established charmonium-(like) states, $\psi(4160),\psi(4230)$, $\psi(4360)$, $\psi(4415)$, and $\psi(4660)$, are checked and no evidence for any $\psi(Y) \to pK^-\bar{\Lambda}+c.c.$ decay is found. To set the upper limits of $\Gamma_{\rm ee}\mathcal{B}(\mathcal{R}\to pK^-\bar{\Lambda}+c.c.)$, the likelihood distributions are constructed as $L(\Gamma_{\rm ee}\mathcal{B})=e^{-0.5\chi^2}$. The upper limits at the 90\% confidence level~(C.L.) is obtained by integrating $L(\Gamma_{\rm ee}\mathcal{B})$ from zero to 90\% of the total curve. The uncertainty associated with the quoted resonance parameters of $\mathcal R$ is studied by sampling its parameters according to its uncertainty, repeating the estimation of upper limits, and taking the width of resulting distribution as this uncertainty. The upper limit is assigned as the nominal value plus the uncertainty due to quoted resonance parameters, as summarized in Table~\ref{tab:upper_limits}

\begin{table}[htbp]
	\centering
	\caption{Fixed masses~($M_{\mathcal R}$) and widths~($\Gamma_{\mathcal R}$) together with the upper limits on the product of the $e^+e^-$ partial width and branching fraction of $\mathcal{R}\to pK^-\bar{\Lambda}+c.c.$, $\Gamma_{ee}\mathcal B(\mathcal R\to pK^-\bar{\Lambda}+c.c.)$ at the 90\% C.L. for different charmonium(-like) state.}
	\begin{tabular}{l|ccc}
		\hline
		\hline
		Resonance & $M_{\mathcal{R}}$~(MeV/$c^2$) &  $\Gamma_{\mathcal{R}}$~(MeV) & $\Gamma_{\rm ee}\mathcal{B}(\mathcal R\to pK^-\bar{\Lambda}+c.c.)$~(eV) \\ \hline
		$\psi(4160)$  & $4191\pm5$ & $70\pm10$ & $<3.0\times10^{-3}$ \\
		$\psi(4230)$  & $4223\pm3$ & $48\pm8$ & $<1.6\times10^{-3}$ \\
		$\psi(4360)$  & $4374\pm4$ & $118\pm12$ & $<3.4\times10^{-3}$\\
		$\psi(4415)$  & $4421\pm4$ & $62\pm20$ & $<3.2\times10^{-3}$ \\
		$\psi(4660)$  & $4630\pm6$ & $72\pm14$ & $<2.8\times10^{-3}$\\
		\hline
	\end{tabular}
	\label{tab:upper_limits}
\end{table}

\begin{figure}[htbp]
	\centering
	\includegraphics[width=\columnwidth]{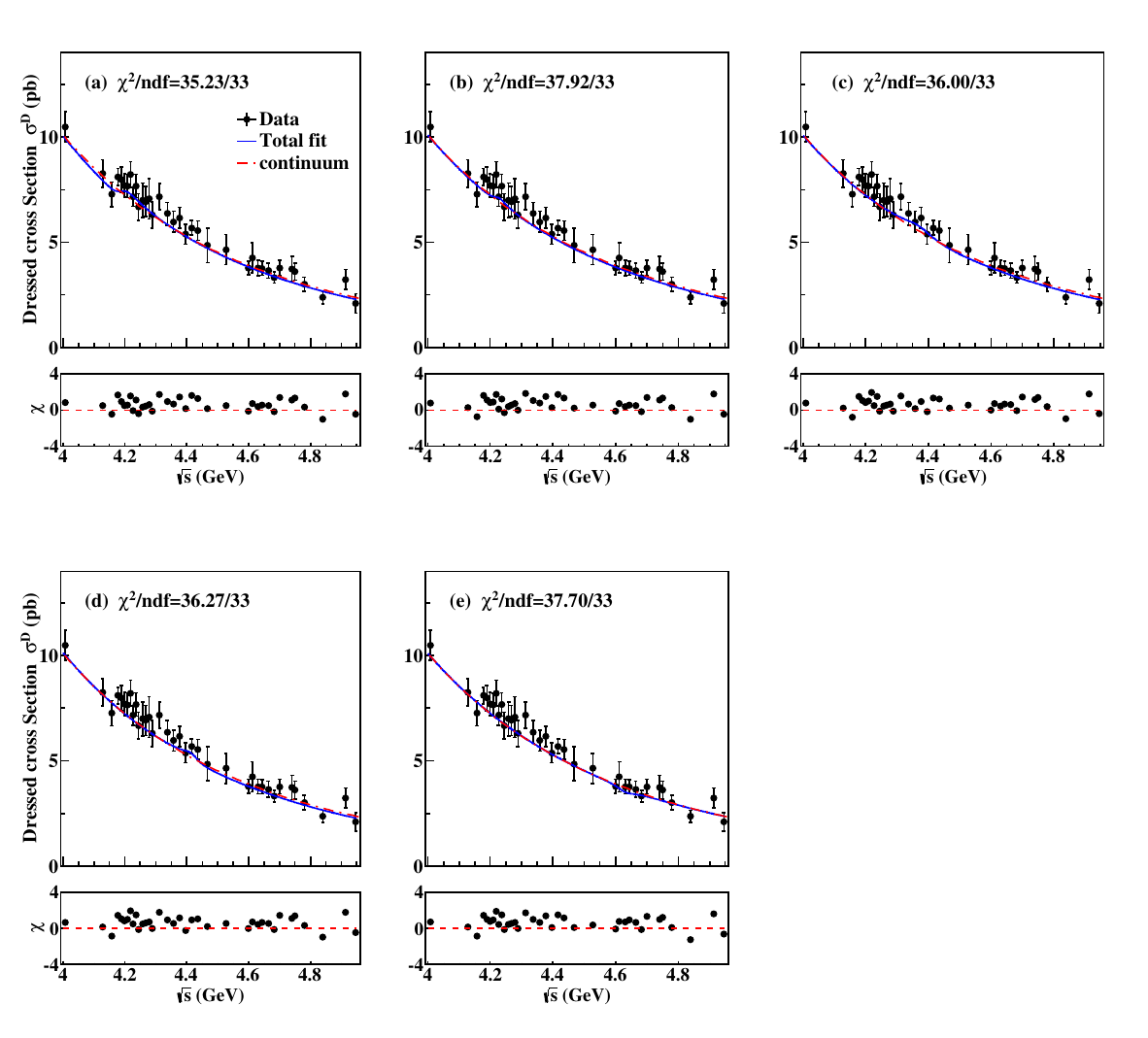}
	\caption{Fits to the energy-dependent dressed cross sections of $e^+e^-\to pK^-\bar \Lambda+c.c.$ under the hypotheses of continuum production plus (a) $\psi(4160)$, (b) $\psi(4230)$, (c) $\psi(4360)$, (d) $\psi(4415)$, and (e) $\psi(4660)$. The points with error bars are the measured values including both statistical and systematic uncertainties. The blue solid curve is the total fit result and the red dash curve is the continuum contribution $a/s^n$.}
	\label{fig:fit_resonance}
\end{figure}

%------------------------------------------------------------------------------
\section{Summary}
\label{chap:SUMMARY}

In summary, with 21.7 fb$^{-1}$ of $e^+e^-$ collision data taken at $\sqrt{s}$ ranging from 4.009 GeV to 4.951 GeV, the energy-dependent cross sections of $e^+e^-\to pK^-\bar{\Lambda}$ are measured for the first time. We fit the obtained cross sections under different hypotheses of charmonium(-like) states plus continuum production. No evidence for any decay of charmonium(-like) states is found and the upper limits of $\Gamma_{\rm ee}\mathcal{B}(\mathcal{R}\to pK^-\bar{\Lambda}+c.c.)$ at the 90\% C.L. are given.

\acknowledgments
%% Saved at => 2023-01-13
The BESIII Collaboration thanks the staff of BEPCII and the IHEP computing center for their strong support. This work is supported in part by National Key R\&D Program of China under Contracts Nos. 2020YFA0406300, 2020YFA0406400; National Natural Science Foundation of China (NSFC) under Contracts Nos. 12175244, 11875170, 11565006, 11505034, 11635010, 11735014, 11835012, 11935015, 11935016, 11935018, 11961141012, 12022510, 12025502, 12035009, 12035013, 12061131003, 12192260, 12192261, 12192262, 12192263, 12192264, 12192265, 12221005, 12225509, 12235017; the Chinese Academy of Sciences (CAS) Large-Scale Scientific Facility Program; the CAS Center for Excellence in Particle Physics (CCEPP); CAS Key Research Program of Frontier Sciences under Contracts Nos. QYZDJ-SSW-SLH003, QYZDJ-SSW-SLH040; 100 Talents Program of CAS; The Institute of Nuclear and Particle Physics (INPAC) and Shanghai Key Laboratory for Particle Physics and Cosmology; ERC under Contract No. 758462; European Union's Horizon 2020 research and innovation programme under Marie Sklodowska-Curie grant agreement under Contract No. 894790; German Research Foundation DFG under Contracts Nos. 443159800, 455635585, Collaborative Research Center CRC 1044, FOR5327, GRK 2149; Istituto Nazionale di Fisica Nucleare, Italy; Ministry of Development of Turkey under Contract No. DPT2006K-120470; National Research Foundation of Korea under Contract No. NRF-2022R1A2C1092335; National Science and Technology fund of Mongolia; National Science Research and Innovation Fund (NSRF) via the Program Management Unit for Human Resources \& Institutional Development, Research and Innovation of Thailand under Contract No. B16F640076; Polish National Science Centre under Contract No. 2019/35/O/ST2/02907; The Swedish Research Council; U. S. Department of Energy under Contract No. DE-FG02-05ER41374.

\bibliographystyle{JHEP}
\bibliography{references.bib}

\end{document}